\documentclass[12pt,preprint]{aastex}

\newcommand{\vt}{V\,773\,Tau}
\newcommand{\vab}{V\,773\,Tau\,AB}
\newcommand{\vc}{V\,773\,Tau\,C}
\newcommand{\vd}{V\,773\,Tau\,D}
\newcommand{\brg}{Br$\gamma$}

\shorttitle{No Fossil Disk in {\vt}}
\shortauthors{Duch\^ene et al.}

\received{2003 January 20}
\begin{document}

\title{No Fossil Disk in the T\,Tauri Multiple System {\vt}}

\author{G. Duch\^ene, A. M. Ghez, C. McCabe}
\affil{Division of Astronomy and Astrophysics, UCLA, Los Angeles, CA
90095-1562}
\email{duchene@astro.ucla.edu}
\author{A. J. Weinberger}
\affil{Department of Terrestrial Magnetism, Carnegie
Institution of Washington,Washington, DC 20015}

\begin{abstract}
  We present new multi-epoch near-infrared (NIR) and optical
  high-angular images of the {\vt} pre-main sequence triple
  system, a weak-line T\,Tauri (WTTS) system in which the
  presence of an evolved, ``fossil'' protoplanetary disk has
  been inferred on the basis of a significant infrared (IR)
  excess. Our images reveal a fourth object bound to the
  system, {\vd}. While it is much fainter than all other
  components at 2\,$\mu$m, it is the brightest source in the
  system at 4.7\,$\mu$m. We also present medium-resolution $K$
  band adaptive optics spectroscopy of this object, which is
  featureless with the exception of a weak {\brg} emission
  line. Based on this spectrum and on the spectral energy
  distribution (SED) of the system, we show that {\vd} is
  another member of the small class of ``infrared companions''
  (IRCs) to T\,Tauri stars (TTS). It is the least luminous,
  and probably the least massive, component of the system, as
  opposed to most other IRCs, which suggests that numerous
  low-luminosity IRCs such as {\vd} may still remain to be
  discovered. Furthermore, it is the source of the strong IR
  excess in the system. We therefore reject the interpretation
  of this excess as the signature of a fossil (or ``passive'')
  disk and further suggest that these systems may be much less
  frequent than previously thought.

  We further show that {\vc} is a variable classical TTS
  (CTTS) and that its motion provides a well constrained
  orbital model. We show that {\vd} can be dynamically stable
  within this quadruple system if its orbit is highly
  inclined. Finally, {\vt} is the first multiple system to
  display such a variety of evolutionary states (WTTS, CTTS,
  IRC), which may be the consequence of the strong star-star
  interactions in this compact quadruple system.
\end{abstract}



\section{Introduction}

{\vt} (HD\,283447, HBC\,367) is a remarkable $\sim5$\,Myr-old,
$\sim1.5\,M_\odot$ (White \& Ghez 2001) pre-main sequence
object located 148\,pc away (Lestrade et al. 1999) in the
Taurus star-forming region. Its SED displays two broad maxima
at $\sim1\,\mu$m and $\sim20\,\mu$m (cf, e.g., data
compilation by Feigelson et al. 1994). The first one
corresponds roughly to what is expected for a stellar
photosphere, while the longer wavelength component is
reminiscent of the spectral energy distribution (SED) of
accreting pre-main sequence stars, i.e. classical T\,Tauri
stars (CTTS). This infrared (IR) excess in CTTS is generally
attributed to a dusty circumstellar disk heated by
reprocessing of starlight and by energy released through the
accretion process. While the IR excess in {\vt} suggests a
disk model for this system, the weak hydrogen emission lines
in its spectrum (e.g., Cabrit et al. 1990) implies that no
accretion is ongoing on the stars, consistent with a weak-line
T\,Tauri star (WTTS), although the presence of forbidden
emission lines in its spectrum imply mass loss. This has been
considered as substantial evidence for the presence of a
``passive disk'' around the stars, i.e., a non-accreting dusty
structure that only reprocesses the stellar
light. Furthermore, Phillips et al. (1996) first pointed out
that, because of the dip around 2\,$\mu$m in the SED of {\vt},
this object presents the characteristics of a ``fossil disk'',
as defined by Strom (1995), in which the inner parts of the
disk have been cleared out of small dust grains possibly
through the formation of planetesimals. In other words, this
system has been proposed to be one of the few ``transition
objects'' between CTTS and WTTS, observed during the phase of
disk dissipation. This duality WTTS/IR excess is present in
only a few objects (e.g., Skrutskie et al. 1990; Stassun et
al. 2001), which suggests that this is a short-lived phase.

While the analysis of the SED and the emission lines have been
based on low angular resolution data sets, high angular and
spectral resolution studies have resolved this object into a
triple system. First, a tight ($\sim0\farcs1\approx15$\,AU)
stellar companion was discovered through speckle imaging
(Ghez, Neugebauer \& Matthews 1993; Leinert et al. 1993). The
primary component of this pair later appeared to be itself
double, a 51\,day-period ($\sim0.3$\,AU semi-major axis; Welty
1995) double-lined spectroscopic binary (SB). All three
components are believed to be WTTS (White \& Ghez
2001). Interactions between the various components are likely
to be important in this tight multiple star system. Indeed,
the variable non-thermal radio emission observed in this
system (O'Neal et al. 1990; Dutrey et al. 1996) appears to be
phase-locked with the SB (Massi, Menten \& Neidh\"ofer
2002). This emission therefore appears to be chromospheric
with enhanced activity occuring when the two stellar
chromospheres partially overlap at periastron. The properties
of the putative fossil disk could also be affected by the
multiplicity of this object. A fit of a disk emission model to
the IR excess places the inner radius of the optically thick
disk at about $\sim0.2$\,AU. This is well inside the orbit of
the SB, raising doubts about the disk model.

In this paper, we present new near-infrared (NIR) and optical
high-angular resolution imaging of {\vt}, which unambiguously
reveal a fourth component. The observed 1.6--4.7\,$\mu$m
colors and the broadband 2\,$\mu$m spectrum of this object
show that it is an infrared comapnion (IRC) and that it is in
fact the source of the IR excess in this system. Throughout
this paper, we will refer to the SB as {\vab}, the long-known
speckle companion as {\vc} and the new companion identified in
this study as {\vd}. We present the various datasets used in
this paper in \S\,\ref{sec:obs-red}.  The orbital motion of
{\vc} and the photometric properties of all components of the
system are analyzed in \S\,\ref{sec:results}. In
\S\,\ref{sec:discus}, we determine the status of each
component and study the importance of star-star interactions
in the system. Finally, we summarize our results in
\S\,\ref{sec:conclu}.


\section{Observations and data analysis}
\label{sec:obs-red}


\subsection{Speckle imaging}

{\vt} has been regularly targeted by speckle interferometry
surveys because, as a $\sim0\farcs1$ binary, it is an
excellent candidate for studies of orbital motion. A number of
measurements (see Table\,1) have already been reported
(Leinert et al. 1993; Ghez et al. 1993, 1995, 1997; Woitas,
K\"ohler \& Leinert 2001, Tamazian et al. 2002).

\begin{figure*}
\plotone{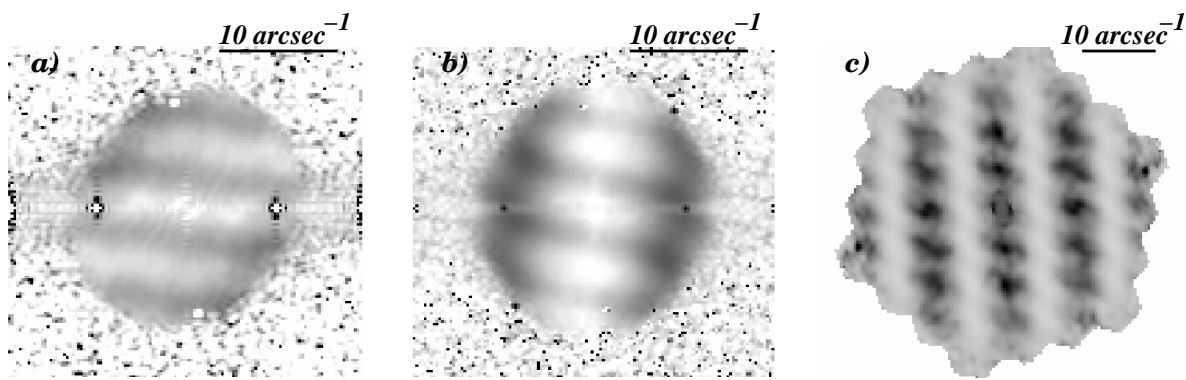}
\caption{Examples of speckle interferometry power spectra of
the {\vt} system obtained in 1997 ($a$), 1999 ($b$) and 2001
($c$). In the first epoch, only the {\vab}--D pair, oriented
more or less North-South is clearly detected. In the later two
epochs, a set of much wider fringes appears, corresponding to
a tight pair oriented roughly East-West, {\vab}--C. In the
2001 dataset, it is particularly clear that two independent
sets of fringes are detected or, equivalently, that the system
is resolved into three different components. Note the
different spatial scales for the Palomar ($a$ and $b$) and
Keck ($c$) datasets, which explains why the fringes associated
to the {\vab}--C pair in the latter show several cycles in the
Fourier space instead of merely one.}
\label{fig:power_spectra}
\end{figure*}

We present here several new NIR speckle observations of the
system. These data were obtained at 2.2\,$\mu$m ($K$ band) on
1996 Oct 24, 1997 Nov 19, 1998 Nov 04, 1999 Nov 20 and 2000
Dec 08 at the 200-inch telescope at Palomar Observatory and on
2001 Dec 08 at the 10\,m Keck I telescope. We used a 64x64
subsection of a 256x256 SRBC InSb aray at Palomar and the
facility instrument Near-IR Camera (Matthews \& Soifer 1994;
Matthews et al.  1996) at Keck. These instruments offer pixel
scales of 0\farcs0337$\pm$0\farcs0005/pixel and
0\farcs0203$\pm$0\farcs0003/pixel respectively, adequately
sampling the diffraction limit in the $K$ band
($\sim0$\farcs09 and $\sim0$\farcs05), and have known
orientations to an accuracy of $\pm$1\degr. Stacks of several
hundred short integration ($t\sim0.1$s) exposures were
obtained, immediately followed by a similar stack of exposures
on a calibration point source. Standard speckle data reduction
routines were applied to the data; we refer the reader to Ghez
et al. (1993) and Patience et al. (1998) for more
details. Examples of power spectra at various epochs are
presented in Figure\,\ref{fig:power_spectra}.

In the latter four epochs, the calibrated power spectra of the
system reveal two distinct sets of fringes, corresponding to
the {\vab}--C and {\vab}C--D pairs\footnote{The second set is
in fact the sum of the {\vab}--D and {\vc}--D almost parallel
sets of fringes, which we treat as a single fringe pattern
corresponding to {\vd} and the photocenter of {\vab}C.}. In
those cases, we first fit the better-defined {\vab}-C pair,
then divide the observed power spectrum of the system by a
model of this binary and finally fit a second binary
system. The information regarding the {\vab}--D pair is then
easily recovered via the known separation and flux ratio of
{\vab}--C. A regular binary system model is fit to the other
datasets, which do not show sufficient evidence for a third
resolved component\footnote{In the 1997 dataset, the {\vab}--C
is not clearly resolved although the amplitude of the power
spectrum decreases towards the outer edge of the power
spectrum along PA$\sim40${\degr} (see
Figure\,\ref{fig:power_spectra}). Since we do not detect the
first minimum of the visibility curve along this direction, we
can only set an upper limit on the binary separation
($\lesssim\lambda/2D\approx45$\,mas in the $K$ band with the
Palomar 200-inch telescope). In subsequent epochs, this first
minimum is reached and the binary parameters can be
confidently derived.}.

We also present here the results of a visible wavelength
speckle interferometry experiment performed on 1991 Sep 25 to
29 at the 200-inch telescope at Palomar Observatory with the
same setup as that described by Gorham et al. (1992). The
filters used, $F_{620}$, $F_{656}$, $F_{800}$ and $F_{850}$,
are all 100\,{\AA} wide and are centered on 6200, 6560, 8000
and 8500\,{\AA} respectively. Data reduction and analysis
routines are also similar to that described by Gorham et
al. (1992), in which autocorrelation functions (ACFs) are
generated. While ACFs are related to the power spectrum
analyzed for the IR data, they are more robustly generated in
photon-counting experiments. Only two stars are identified in
our data, which are readily associated with {\vab} and C.


\subsection{Adaptive optics imaging}

\subsubsection{$H$ \& $K$ band imaging}

\begin{figure}
\epsscale{0.95}
\plotone{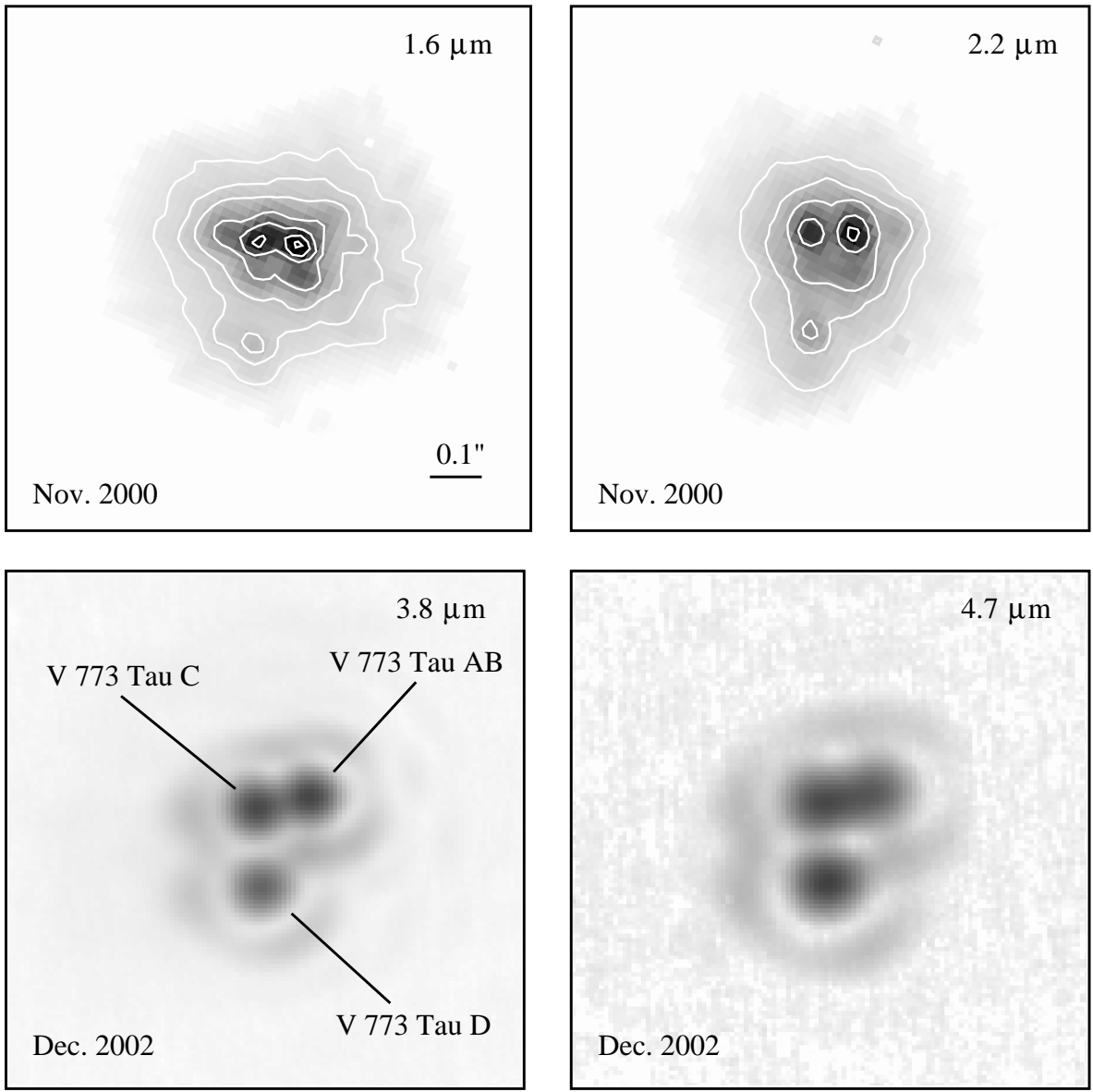}
\caption{Adaptive optics images of the {\vt} multiple system
obtained in November 2000 (top row) and December 2002 (bottom
row). A square root stretch has been used in all images. North
is up and East to the left and each image is 1$\arcsec$
across. The lowest contour in the $H$ and $K$ images is about
120 times and 150 times higher than the background noise. The
contours represent 2\,\%, 3\,\%, 5\,\%, 15\,\%, 33\,\%, 60\,\%
and 90\,\% of the peak value in the $H$ band image and 3\,\%,
7\,\%, 15\,\%, 45\,\% and 90\,\% of the peak at $K$. The
motion of {\vc} with respect to {\vab} between the two epochs
is particularly obvious in these images, as well as the
relative brightening of {\vd} towards longer wavelengths.}
\label{fig:im_ao}
\end{figure}

In addition to this wealth of speckle measurements, we also
obtained $H$ band\footnote{The filter we used, ``NIRSPEC-3'',
is a customized filter specific to the instrument. Its central
wavelength ($\lambda_c=1.625\,\mu$m) and bandwith
($\Delta\lambda=0.36\,\mu$m) are close enough to the standard
broadband $H$ filter ($\lambda_c=1.630\,\mu$m,
$\Delta\lambda=0.31\,\mu$m) that we do not expect magnitude
differences larger than a few hundredths. Throughout our
paper, we therefore refer to this filter as the $H$ filter.} 
and $K$ band adaptive optics (AO) images of the {\vt} system
with the AO system of the 10\,m Keck II telescope (Wizinowich
et al. 2000) and the facility NIR spectrometer NIRSPEC (McLean
et al. 2000). The data were obtained on 2000 Nov 20 with SCAM,
the slit-viewing camera inside NIRSPEC. With each filter, we
obtained 4 sets of 100 individual 0.1\,s exposures, with the
object centered on a different part of the detector in each
set. The individual exposure time was short enough to avoid
saturating the stellar cores. {\vt} is bright enough in the
visible ($V\sim10.7$, $R\sim9.8$) and is unresolved in the
wavefront sensor camera so it could be used as a natural guide
star for the AO system. We achieved Strehl ratios of 12\,\%
and 15\,\% respectively at $H$ and $K$, which is enough to
provide diffraction-limited cores. The resolution of the
images, measured as the FWHM of a Gaussian fit to the radial
profile of the brightest component, {\vab}, are about 40 and
50\,mas at $H$ and $K$, respectively. The final images are
presented in Figure\,\ref{fig:im_ao} (upper row).

Data reduction included sky subtraction, flat-fielding, bad
pixel removal and shift-and-adding the individual images and
was performed in IRAF\footnote{IRAF is distributed by the
National Optical Astronomy Observatories, which is operated by
the Association of Universities for Research in Astronomy,
Inc., under contract to the National Science
Foundation.}. Astrometry and relative photometry are performed
through two methods. First, we estimate the centroid locations
for each component in the final image. To estimate
uncertainties, we perform the same analysis on four subsets of
the data and calculate their standard deviation. Relative
photometry is obtained by summing fluxes over 2-pixel radius
apertures around each component; uncertainties are similarly
obtained through subsets of the data. To avoid potential
star-to-star flux contamination, we also fit point spread
functions (PSFs) to the multiple system. Because of the
limited field of view of SCAM, no other star is available to
be used as a contemporaneous PSF. We thus use both a
theoretical image which is basically the Fourier transform of
the entrance pupil of the telescope and an empirical PSF with
a somewhat better image quality, which we obtained on Nov 19
by imaging T\,Tau\,N (see Duch\^ene, Ghez \& McCabe 2002 for
details). All three techniques agree within 3\,$\sigma$ of
each other and all trends are naturally explained by the
different imperfections in the PSFs. The final astrometric and
photometric results presented in Table\,1 are the averages of
these three techniques (one direct measurement and two fits).

\subsubsection{$L'$ \& $M'$ band imaging}

We also imaged {vt} at 3.8\,$\mu$m and 4.7\,$\mu$m ($L'$ and
$M'$ broadband filters, which are 0.7\,$\mu$m and 0.24\,$\mu$m
wide respectively) with the new facility instrument NIRC2
(Matthews et al., in prep.) behind the AO system on the Keck
II telescope. The data were obtained on 2002 December 13 under
excellent seeing conditions; {\vt} itself was used as a
natural AO guide star. Strehl ratios on order of 60-75\,\%
were measured at $L'$ on similarly bright single stars
throughout the night. A circular cold mask equivalent to a
circular 9\,m-diameter entrance pupil was used to minimize the
background emission, a critical factor in the thermal IR
range. The resulting diffraction-limited images have Gaussian
FWHM of 90 and 110\,mas at $L'$ and $M'$, respectively.

Several hundred short exposures ($\sim0$.05\,s each) were
obtained at four different locations on the detector for a
total integration time of about 1\,min with each filter. The
datasets were subtracted from each other to remove the thermal
background, flat-fielded, corrected for bad pixels and
shift-and-added to produce the final images presented in
Figure\,\ref{fig:im_ao} (lower row). Astrometric and
photometric parameters of the multiple systems were obtained
using the same methods as described above for the $H$ and $K$
images. The PSF fitting process is performed using HD\,1160 (a
single star) at $L'$ and T\,Tau\,N at $L'$ and $M'$. Both were
observed on the same night with the same setup. All methods
yield compatible estimates within $\sim2\,\sigma$ for both the
location of the centroids and the flux ratios.


\subsection{Adaptive optics spectroscopy}

On 2000 Nov 20, we also obtained long-slit $K$ band
spectroscopy of two components of the system, by aligning the
$0\farcs036$-wide slit of NIRSPEC along the {\vab}--{\vd}
position angle. Four spectra were obtained at different
locations behind the slit, with a total integration time of 10
minutes. The instrumental setup is identical to that used in
our previous study of the tight binary T\,Tau\,S reported in
Duch\^ene et al. (2002), which also provides more details on
the data reduction process. The spectra cover the
2.0--2.4\,$\mu$m range, with a 2-pixel resolution of
$R\sim2500$. Immediately after {\vt}, we observed the nearby
stars HD\,27311 (A0) and HD\,27560 (G0) to correct for
telluric absorption features. The spectra were first
sky-subtracted, corrected for distortion effects, flat-fielded
and bad pixel-corrected.

To extract the individual spectra, we performed a fit to the
binary profile along the spatial axis with a semi-empirical
PSF. The latter is obtained by unfolding the Northern half of
the profile of {\vab}, i.e., assuming a symmetric profile. The
same profile was fitted to both components simultaneously. We
estimate that the contamination of the spectrum of {\vd} by
that of {\vab} is less than 5\,\% of the former at any
wavelength. The continuum around {\brg} in the spectra of
{\vt} was not perfectly corrected with HD\,27560, either
because its spectrum is not close enough to the solar spectrum
or because the telluric features changed between our
observations. We thus divided both spectra by a third-order
polynomial function fit to a 40\,nm-wide window centered on
{\brg} and containing no other photospheric feature. The
resulting spectrum of {\vab} in this window (except for about
ten pixels around the hydrogen line itself) was used to
correct the spectrum of {\vd} of any remaining telluric
feature. This procedure does not modify the shape or strength
of the {\brg} line in either spectrum but it improves
significantly the continuum around that line. Finally, all
spectra were divided by a fifth-order polynomial function fit
to their respective continua. The output spectra are presented
in Figure\,\ref{fig:spectra}. The signal-to-noise ratio is
estimated to be on the order of 100 per pixel in the final
spectra.


\section{Results}
\label{sec:results}


\subsection{A new stable quadruple system in Taurus}
\label{subsec:quad}

As can be seen in Figure\,\ref{fig:im_ao}, the system is
clearly resolved in all our AO images into three components at
a spatial resolution of 0\farcs05--0\farcs1. The brightest
component at 1.6--3.8\,$\mu$m also dominates the flux in the
visible and is associated with {\vab}, the SB discovered by
Welty (1995). The latter cannot be resolved in our data; even
at apoastron its projected separation on the sky is
0\farcs0029. The three visual components are also
simultaneously detected in the latest epochs of speckle
imaging, from 1998 onward (see
Figure\,\ref{fig:power_spectra}). We attempted to re-analyze
earlier datasets, but the fainter component did not show up
because of the lower quality of earlier IR detectors,
variability of the source and/or a less favorable
configuration of the system. Our multiwavelength dataset
unambiguously reveals that {\vab} has two distinct apparent
companions.

\begin{figure}
\plotone{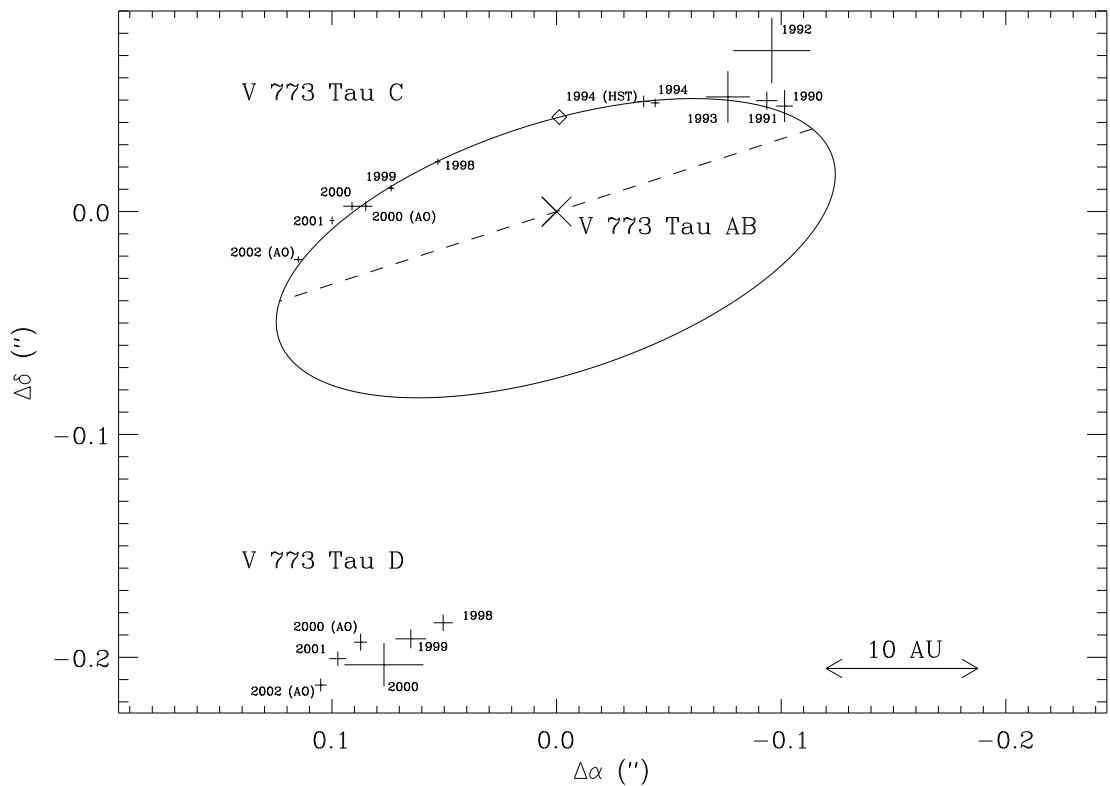}
\caption{Location of the companions that have been spatially
resolved from {\vab} since 1990. The SB (large cross) is fixed
at location (0,0); the amplitude of the internal orbital
motion of {\vab} is negligible at the scale of this
figure. The length of the tick marks indicate 1$\,\sigma$
error bars and the year of observation is associated to each
point. Only the datapoints used in the dynamical analysis are
shown here. Our best fit orbit for {\vc} is represented by the
solid curve, together with its periastron (open diamond) and
line of nodes (dashed line).}
\label{fig:astrom}
\end{figure}

By combining our new astrometric results with previous
measurements, we can study the relative motion of these stars.
We choose to use the rest frame of {\vab}. We exclude the 1991
measurement of Leinert et al. (1993) and the 1993 $J$ band
measurement of Woitas et al. (2001) from our analysis. The
former results are obtained from deprojecting the visibilities
measured along two orthogonal axes while the latter are
obtained with an instrument setup that does not provide
Nyquist sampling of the PSF thus creating
aliasing. Furthermore, to minimize the number of instrumental
set-ups used in this analysis without reducing the time
coverage of the orbit, we do not include the measurements
presented in Tamazian et al. (2002). The locations of the
companions as a function of time are illustrated in
Figure\,\ref{fig:astrom}. The motion of both visual companions
to {\vab} are almost linear over the truncated 4 year time
baseline 1998--2002, with velocities of 18.0$\pm$0.4\,mas/yr
at position angle 125\fdg1$\pm$2\fdg1 and 13.7$\pm$1.1\,mas/yr
along position angle 120\degr$\pm$8{\degr} for {\vc} and {\vd}
respectively. While the proper motion of the system is of
similar amplitude, its position angle is different
($\sim152\degr$, Frink et al. 1997), and it is unlikely that
{\vd} is a foreground/background object unrelated to the other
components of {\vt}. This is further reinforced by the very
small separation of the {\vab}--D pair. The surface density of
$K\leq9.0$ objects (since $K_{AB}\sim7.0$, Ghez, White \&
Simon 1997, we estimate that $K_D\sim9.0$) around {\vt} in the
2 Micron All-Sky Survey is about $2.5\times10^{-6}$ per square
arcsecond. The probability of finding such a
foreground/background object within $0\farcs25$ of {\vab} is
only $\approx5\times10^{-7}$. We therefore conclude that {\vd}
really is physically associated with {\vt}, which is therefore
a quadruple system.

With measurements regularly sampling the last twelve years,
the orbit of {\vc} is well defined. We therefore fit a
Keplerian orbit to our astrometric dataset to constrain the
orbital parameters, in particular the total system mass. We
neglect here the influence of {\vd} on the {\vab}--C
system. We assumed that the orbit is closed ($e<1$), fixed the
distance to $D=148$\,pc, and explored the 7-dimension
parameter space ($P$, $a$, $e$, $i$, $\omega$, $\Omega$,
$T_0$) using a $\chi^2$ minimization routine. We used 10000
different starting points that were randomly selected within
the allowed range for each orbital parameter to make sure that
we converge on the absolute minimum of $\chi^2$. The resulting
best fit orbit ($\chi^2=1.94$) is shown in
Figure\,\ref{fig:astrom} and the corresponding orbital
parameters are presented in Table\,2. The estimated
inclination is very similar to the orbit of the SB (Welty
1995), suggesting a coplanar triple system. The derived
periastron and apoastron distances are $14\pm1$\,AU and
$26\pm4$\,AU respectively. Until relative radial velocities
within the system are measured, an ambiguity on the
inclination cannot be solved and, independently of this, the
total system mass is only determined to within a
distance-related scaling factor: $M_{ABC} = (3.7\pm0.7) \times
(D/148\,{\rm pc})^3\,M_\odot$. Therefore, the uncertainty on
the distance to the system (Lestrade et al. 1999) introduces
an additional 10\,\% uncertainty on the system
mass. Surprizingly, the uncertainties we estimate on the
orbital parameters are not smaller than those derived by
Tamazian et al. (2002) in spite of our extended coverage of
the orbit. Some parameters differ by up to 7--9\,$\sigma$
($P_{orb}$, $a$) between the two orbital solutions, suggesting
that Tamazian et al. underestimated their uncertainties.

The subsystem {\vab}C is a hierarchical triple system with a
ratio of orbital periods of $P_{AB-C}/P_{A-B}\gtrsim300$; it
is therefore expected to be dynamically stable over long
timescales. On the other hand, the projected separation of
{\vd} is only slightly larger than the apoastron distance of
{\vc}, which raises the question of the stability of its
orbit. Since {\vab} has such a compact orbit, we can consider
it a single point mass. We apply the following criterion,
derived by Eggleton \& Kiseleva (1995) for a non-coplanar
triple system to {\vab}--C--D: for a given combination of the
mass ratios in a hierarchical system, there is a critical
value of the ratio of the periastron distance of the outer
pair to the apoastron separation of the inner binary beyond
which the system can be considered stable for at least several
hundred times the outer period. If this criterion is not
fulfilled, there are times at which the three star-to-star
distances in the system are commensurable, which is a highly
unstable configuration. Applying this criterion with $q_{in} =
M_{AB}/M_C \approx 3.7$ and $q_{out} = M_{ABC}/M_D \gtrsim
3.1$ (see \S\,\ref{subsec:nature}), we find that the orbit of
{\vd} is stable if its periastron distance is larger than
about 115\,AU. Since its projected separation is about 30\,AU,
this suggests that its orbit is highly inclined to the line of
sight ($i\gtrsim75$\degr, similar to the $\sim67$\degr\
inclination of both {\vab} and {\vc}). If {\vd} were on a
circular orbit with such a radius, its predicted orbital
velocity would be $\sim6$\,km/s. This is very similar to its
observed projected velocity with respect to the center of mass
of {\vab}--C, $V_D=6.9\pm0.8$\,km/s as measured from a linear
fit to the motion of {\vd} over the 1998-2002 period. Our
astrometric dataset is therefore consistent with {\vd} being
on a stable, inclined orbit around the inner triple system.


\subsection{SED and variability of the components of {\vt}}
\label{subsec:sed}

\begin{figure}
\plotone{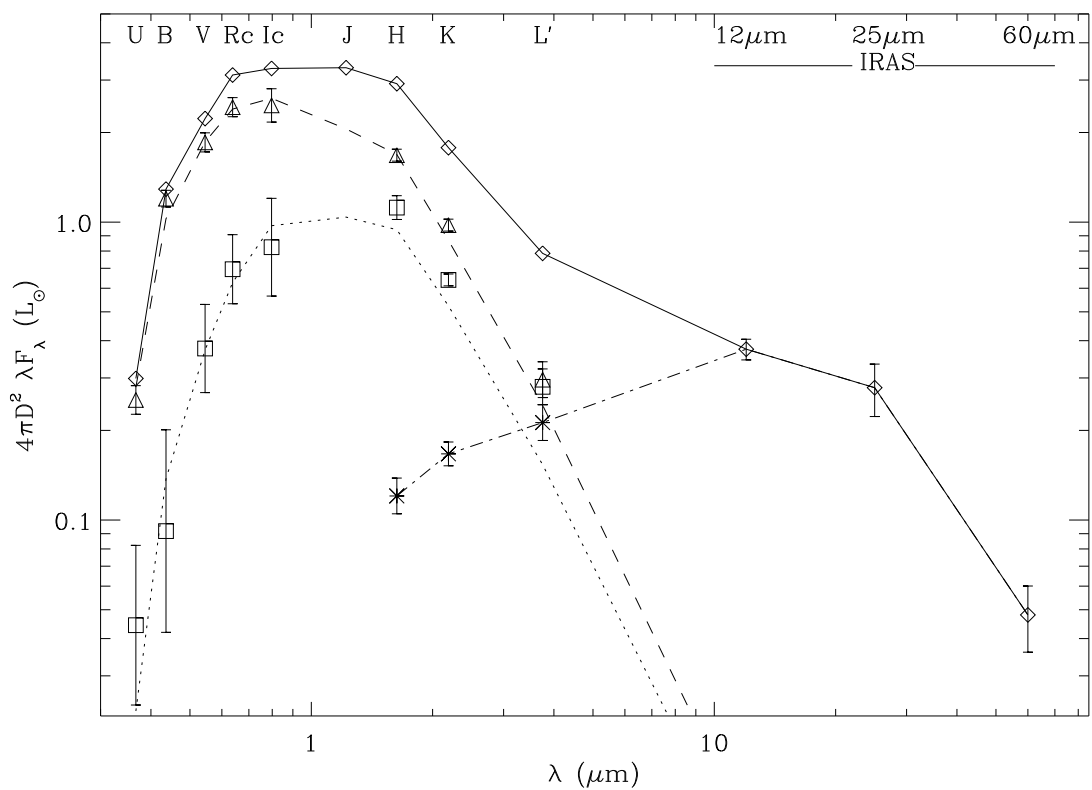}
\caption{Dereddened SED of the components of the system. {\it
Diamonds}: unresolved system; {\it triangles}: {\vab}; {\it
squares}: {\vc}; {\it asterisks}: {\vd}.  The dashed curve is
the sum of the K2 and K7 photospheres fitted to the components
of {\vab} while the dotted curve is the M0.5 photosphere fit
to {\vc}. The dot-dashed curves joins the observed $HKL'$
photometric points for {\vd} with the unresolved $IRAS$
measurements.}
\label{fig:sed}
\end{figure}

In an effort to understand the nature of {\vd}, we first
construct the SED of the system from 4000\,{\AA} up to
60\,$\mu$m. Unresolved photometry for the quadruple system is
taken from White \& Ghez (2001) in the visible, Rydgren \&
Vrba (1981) in the 1--4\,$\mu$m range and Prusti et al. (1992)
for $IRAS$ measurements\footnote{The $IRAS$ observations are
blended with the CTTS FM\,Tau, which is located 40{\arcsec}
away. Prusti et al. (1992) were able to disentangle the
contributions of both sources at 12 and 25\,$\mu$m. At
60$\,\mu$m, they assumed the same flux ratio as observed at
25\,$\mu$m. This extraction process results in large error
bars (15\,\% and 20\,\% at 25 and 60\,$\mu$m
respectively).}. Flux ratios within the system are from White
\& Ghez (2001) in the visible and from this study (our AO
measurements) from 1.6\,$\mu$m to 3.8\,$\mu$m. No absolute
photometric measurement of {\vt} is available at 4.7\,$\mu$m,
so we did not use the results of our $M'$ image in this
analysis. Finally, we used the $R$ band flux ratio of the SB
from Welty (1995) to separate the contributions of {\vt}\,A
and B. To correct for the foreground interstellar extinction,
all measurements were dereddened with $A_V=1.39$\,mag (Welty
1995, White \& Ghez 2001) using a standard extinction law
(Rieke \& Lebofsky 1985). We assumed that none of the
components of {\vt} was affected by circumstellar
extinction. The resulting SED is presented in
Figure\,\ref{fig:sed}.

\begin{figure}
\plotone{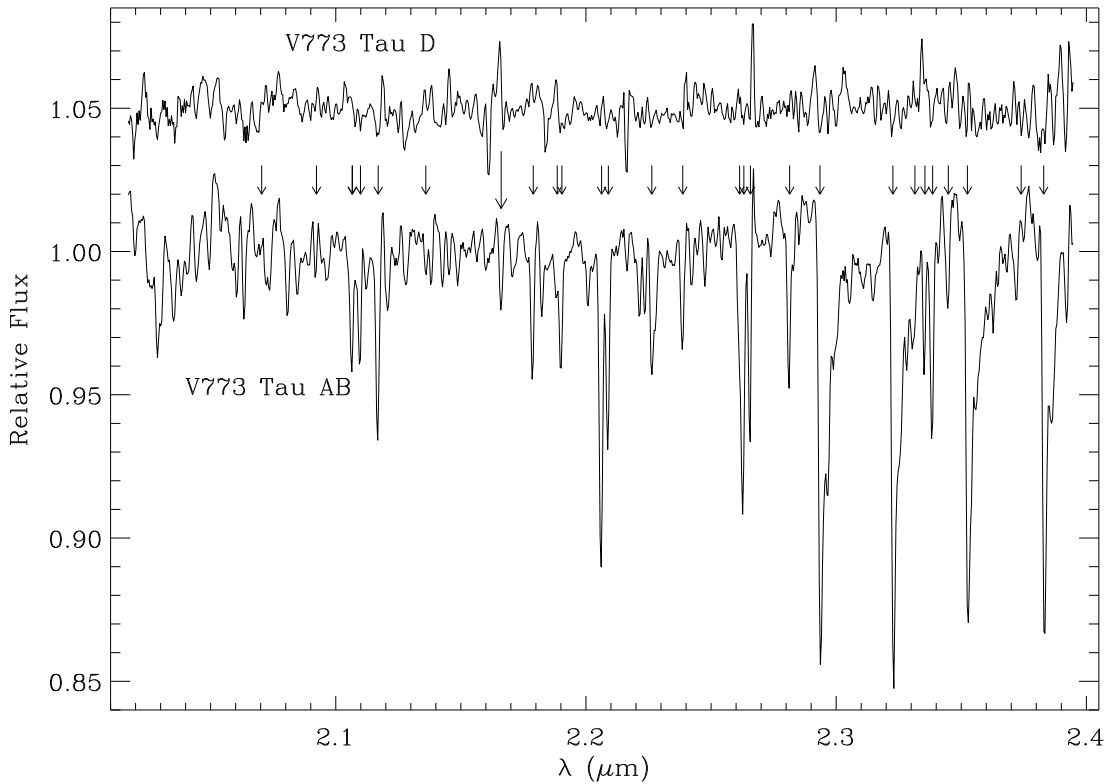}
\caption{Continuum-normalized spectra of {\vab} and {\vd}; the
latter has been shifted upwards by 0.05 for clarity. The arrows
indicate all known spectral features in this wavelength range
(Kleinmann \& Hall 1986) and the {\brg} line is indicated by
the longer arrow. Both spectra have been smoothed down to a
resolution of about $R\sim1900$. A few spikes, both in
emission and in absorption, appear in the spectrum of {\vd}
but they are too narrow (typically 1 pixel-wide) to be real
features. Furthermore, similar ``features'' can be seen in the
spectrum of {\vab} (e.g., at 2.217\,$\mu$m or 2.267\,$\mu$m),
which suggests that they correspond to poorly corrected
telluric absorption lines.}
\label{fig:spectra}
\end{figure}

We fit photospheric SEDs (as tabulated by Kenyon \& Hartmann
1995) to the two components of the SB using the $VRI$
photometry since this wavelength range is the least likely to
be affected by accretion-induced excesses that can be present
in TTS. Initially, the spectral types are considered to be
those derived in the literature, K2 and K5 (Welty 1995), but a
marginally better fit is obtained with spectral types K2 and
K7 for {\vt}\,A and B respectively; the associated stellar
luminosities are 2.2 and 1.4\,$L_\odot$, with an uncertainty
of about 0.1\,$L_\odot$. White \& Ghez (2001) obtained a
similar value for the luminosity of {\vt}\,A although they
estimated it based on the $I$ band photometry only. We do not
find evidence for any excess above photospheric levels up to
3.8\,$\mu$m for this pair. The unresolved spectrum of this
pair displays many photospheric features, including a weak
{\brg} absorption line (see Figure\,\ref{fig:spectra}). From a
comparison of this spectrum with published spectra libraries
(Kleinmann \& Hall 1986; Wallace \& Hinkle 1997), we estimate
that the strength of the various features is compatible with a
spectral type in the range K0--K5IV, consistent with the
spectral types adopted here.

As already pointed out by White \& Ghez (2001), the SED of
{\vc} can be fit with a pure photosphere from $U$ to
$K$. However, a $\sim3\,\sigma$ excess is then present at
3.8$\,\mu$m, which is further reinforced by the fact that
{\vc} becomes brighter than {\vab} at $M'$. The nature of
{\vc} is revisited in \S\,\ref{subsec:nature}, where we argue
that it is in fact a CTTS. For now, we only fit a photosphere
to the visible measurements, and find a best fit for a
spectral type M0.5 with a 1 subclass uncertainty and a
$1.5\pm0.2\,L_\odot$ stellar luminosity, also in agreement
with those obtained by White \& Ghez (2001). This results in a
$\sim0.2$\,mag excesses at both $H$ and $K$, a $\sim0.7$\,mag
excess at $L'$ and a $K-L' \approx 0.65 \pm 0.15$\,mag color
index. Broadly speaking, SED fitting is a trade-off between
spectral type, luminosity and line-of-sight extinction. The
assumption of identical extinction toward {\vc} and {\vab} can
be relaxed, in which case our estimated spectral type and
luminosity are modified. With an additional 1\,mag of
extinction at $V$ toward {\vc}, we find that its SED is well
fit by a K7 dwarf with a luminosity of $2.3\pm0.3\,L_\odot$; a
significant excess at $L'$ remains present although the $H$
and $K$ band fluxes are then consistent with photospheric
levels. Larger extinctions are not allowed, since it would
imply that more than $0.8\,L_\odot$ is absorbed by dust and
reemitted at longer wavelengths, while this is the total
observed mid- to far-infrared luminosity in the system (see
below). We adopt here the ``equal-$A_V$'' estimates for {\vc},
i.e. a M0.5 spectral type and a 1.5$\,L_\odot$ luminosity.

In contrast to these three components, which can be fitted
with normal photospheric colors and a moderate NIR excess for
{\vc}, {\vd} displays extremely red colors ($H-K\sim1.3$\,mag,
$K-L'\sim1.7$\,mag). A blackbody fit to its dereddenned
$H,K,L'$ photometric data yields $T_{NIR}=900\pm200$\,K, which
is much cooler than any stellar photosphere. Its
1.6--3.8\,$\mu$m SED is rising and this extends to longer
wavelengths since {\vd} is the brightest component of the
system in our $M'$ image, suggesting that it might the
dominant source of the unresolved mid-infrared flux of the
system.  Assuming that the $IRAS$ fluxes can be fully assigned
to {\vd}, we conclude that the SED of this object presents a
broad peak from $\sim5\,\mu$m to $\sim30\,\mu$m and that its
bolometric luminosity is $0.8\pm0.1\,L_\odot$. Since {\vc} may
exhibit a significant excess beyond the $M'$ band, either
through reprocessing of absorbed visible starlight or through
emission of its disk, this value is an upper limit to the
luminosity of {\vd}. Nonetheless, we consider that the
extremely red 1.6--4.7\,$\mu$m colors of {\vd} are substantial
evidence that the mid-IR emission arises predominantly from
this component. We also find that the spectrum of {\vd} is
remarkable in that it shows no photospheric features, in spite
of the high signal-to-noise ratio of our data
(Figure\,\ref{fig:spectra}). The only significant feature in
this spectrum is the {\brg} line, which displays a weak
emission with an equivalent width of $0.4\pm0.1$\,\AA. Since
all photospheres have identifiable features in the observed
wavelength range, we conclude that the NIR light we are
receiving from {\vd} has been thermally reprocessed by dust.

Finally, we note that at least some of the components in this
system are photometrically variable, potentially increasing
the uncertainties on the luminosities derived above. For
instance, the flux ratio between {\vab} and {\vc} varies by
1--1.5\,mag both in the visible and the NIR (see Table\,1). In
fact, the resolved photometry presented by Ghez et al. (1997)
suggests that {\vc} is the variable component, as its $K$ band
magnitude varied by $\sim1$\,mag while {\vab} remained almost
unchanged. Furthermore, {\vab}, which dominates the flux of
the system in the visible, has been shown to vary by only
$\sim0.15$\,mag in the visible (Rydgren \& Vrba
1983). Similarly, the {\vab} to {\vd} flux ratio at $K$ band
varied by $\sim1$\,mag between 1998 and 2002 implying that
{\vd} is also variable. In the absence of simultaneous visible
and NIR photometry for all components of the system, it is not
possible to decide whether the variability of {\vc} and {\vd}
is due to varying extinction along the line of sight or is
intrinsic to these objects.


\section{Discussion}
\label{sec:discus}


\subsection{Nature of the components of the system}
\label{subsec:nature}

\subsubsection{{\vab} and {\vc}}

{\vab} has so far been studied spectroscopically at optical
wavelengths, under the assumption that the contamination by
{\vc} is negligible. The weakness of the H$\alpha$ emission
line in its spectrum (2--4\,{\AA} equivalent width; Herbig,
Vrba \& Rydgren 1986; Cabrit et al. 1990; Feigelson et
al. 1994) as well as the chromospheric-like ultraviolet Mg\,II
$h$ and $k$ emission lines (Feigelson et al. 1994), have led
to a WTTS classification of both components of the SB. The
observed {\brg} line in absorption, the absence of NIR
broadband excess in its SED up to $L'$, and the very limited
amount of visible and NIR variability (\S\,\ref{subsec:sed})
are also consistent with this classification, and we therefore
adopt it. Based on their estimated spectral types and
luminosities, White \& Ghez (2001) concluded that
$M_A=1.46\,M_\odot$. Our adopted luminosity is marginally
larger, and we adopt $M_A \approx 1.5\pm0.1\,M_\odot$. Using
the dynamical mass ratio estimated by Welty (1995) for the SB,
this implies a mass of $\approx1.1\,M_\odot$ for {\vt}\,B.

The classification of {\vc} as a WTTS by White \& Ghez (2001)
was based on its photopheric-like SED. However, as pointed out
in \S\,\ref{subsec:sed}, it displays conclusive excesses at
$L'$ and $M'$ as well as possible excesses at $H$ and $K$. Its
NIR excess is typical of CTTS and unheard of in
WTTS. Similarly, the observed variability of {\vc} in the
visible and the NIR is typical of CTTS: variations of more
than $\sim0.5$\,mag in the visible and/or the NIR are
generally associated with actively accreting TTS (Herbst et
al. 1994; Skrutskie et al. 1996). It thus appears that {\vc}
is a CTTS, and such a classification resolves some of the
puzzling properties of this system. These include the broad
H$\alpha$ emission line found by Feigelson et al. (1994), the
strong {\brg} emission line\footnote{In a 1.3$\,\mu$m spectrum
obtained two nights earlier, they failed to detect the
Pa\,$\beta$ line in emission, even though it is usually
stronger than the {\brg} in CTTS (Muzerolle et al. 1998). This
probably indicates that the emission lines are strongly
variable.} measured by Folha \& Emerson (2001) and the
detection by Carbit et al. (1990) of visible forbidden
emission lines, which are all atypical for a WTTS and rather
suggest CTTS-like ongoing accretion and mass loss.  All of
these features can be explained if {\vc} is a normal CTTS
whose spectrum has been diluted by the spectrum of its bright
WTTS neighbour since it accounts for only 20--30\,\% of the
flux of the unresolved system. We therefore conclude that
{\vc} is a CTTS, which should be confirmed in the future by
obtaining separate spectra for {\vab} and {\vc} in the visible
or the NIR.

Estimating the mass of {\vc} is a non-trivial task, as it lies
above the 1\,Myr isochrone of the evolutionary model of
Baraffe et al. (1998), with either M0.5 or K7 adopted as its
spectral type, while {\vt}\,A falls on the 5\,Myr isochrone
(White \& Ghez 2001). This is most likely because the
luminosity of {\vc} is partially contaminated by
accretion-induced emission and is not purely
photospheric. White \& Ghez estimated an M0 spectral type and
$M_C=0.7\,M_\odot$ because the object lies in the HR diagram
just above this almost vertical evolutionary track. Applying a
similar reasoning, our K7--M0.5 estimated spectral type
implies a stellar mass in the range 0.65--0.8$\,M_\odot$. We
therefore adopt $M_C=0.7\pm0.1\,M_\odot$. The total mass of
the {\vab}C triple system is therefore predicted to be
$3.3\pm0.2\,M_\odot$, which is in agreement with the dynamical
mass we derived in \S\,\ref{subsec:quad}. While a finer
analysis of the stellar masses in the system will require
radial velocity measurements, this indicates that the stellar
masses we adopt here are coherent with our dynamical analysis.

\subsubsection{{\vd} is an ``infrared companion''}

As noted in \S\,\ref{subsec:sed}, the SED of {\vd} cannot be
fit with any stellar photosphere and its spectrum suggests
that this object is seen through optically thick
dust. Therefore, it cannot be classified as a usual TTS.  In
fact, the shape of its SED matches that of a handful of
objects first identified as IRCs to TTS by Chelli et
al. (1988) and extensively studied by Koresko, Herbst \&
Leinert (1997). These companions were initially defined as
extremely red, cool, objects located within a few hundred AU
of a TTS. In agreement with our observations of {\vd}, these
IRCs have SEDs that are too broad to be fitted by single
blackbody curves and that dominate the flux from the systems
at wavelengths longer than 2--5\,$\mu$m. Furthermore, $K$ band
spectroscopy of several other IRCs have revealed featureless
spectra, with the exception of {\brg} in emission (Beck, Prato
\& Simon 2001; Duch\^ene et al. 2002; Herbst, Korseko \&
Leinert 1995). Overall, we consider that the available dataset
regarding {\vd} clearly supports its classification as a new
member of the small category of IRCs.

Although {\vd} shares enough properties with other IRCs to be
classified as such, it stands out from the other members of
this class in at least two respects. The first remarkable
property of {\vd} is that its bolometric luminosity is not
larger than that of the system's primary. In fact, all other
components in the system are intrinsically brighter than the
IRC. The upper limit on its bolometric luminosity places a
strict upper limit on its mass of about 1.2\,$\,M_\odot$
(Baraffe et al. 1998, Palla \& Stahler 1999, Siess et
al. 2000), possibly significantly less if its luminosity is
partially driven by accretion. Following the analysis of
Do-Ar\,24E by Koresko et al. (1997), we believe that the low
luminosity of {\vd} indicates that its mass is smaller than
that of the other components in the system. We therefore
conclude that $M_D \lesssim 0.7\,M_\odot$.

The other noticeable property of {\vd} can be found in the
8--13\,$\mu$m spectrum of the unresolved system presented by
Hanner, Brooke \& Tokunaga (1998), which displays no silicate
feature, either in absorption or in emission. If the
mid-infrared flux of {\vt} is dominated by the IRC, this
contrasts with all other IRCs for which a spectrum has been
obtained (Hanner et al. 1998; Natta, Meyer \& Beckwith 2000)
since they all present a strong, broad, spectral feature
around 10\,$\mu$m. The absence of a silicate feature in the
spectrum of {\vd} is unlikely to be the result of a strong
silicate depletion in its envelope, since this has not been
observed among TTS. Possible explanations for this phenomenon
include time variability of the feature, such as observed in
the case of another IRC (Glass\,I, G\"urtler et al. 1999), an
accidental though almost exact cancellation of emission and
absorption features in the envelope, the growth of most
silicate grains beyond a few microns in size, or an unusually
low temperature in the outer parts of the envelope. In cold
environments ($T\lesssim100$\,K), water ice can condense on
and shield small silicates grains, in which case the grain
opacity in the mid-infrared does not show the usual marked
features at 10\,$\mu$m and 20\,$\mu$m (e.g., Pollack et
al. 1994). It is reasonable to expect that the outer parts of
the dusty envelope surrounding {\vd} are colder than that of
other IRCs as it is a less luminous source, though the actual
cause of the atypical mid-infrared spectrum of this source
remains an open issue.

\subsubsection{Nature of the IRCs}

The exact nature of IRCs is still debated. They may be normal
TTS undergoing episodes of intense accretion and which are
embedded in optically thick dusty envelopes with radii as
small as a few AU (Ghez et al. 1991; Koresko et al. 1997;
White \& Ghez 2001; Duch\^ene et al. 2002). They also may be
embedded protostars, i.e. objects in an earlier evolutionary
phase (e.g., Ressler \& Barsony 2001). This latter
interpretation would imply that IRC binary systems are not
coeval, in contradiction with results on large samples of TTS
binaries (Hartigan, Strom \& Strom 1994; White \& Ghez
2001). We emphasize that the observational distinction between
these two types of objects is quite limited; both are highly
embedded sources whose flux is dominated by accretion onto
and/or contraction of the central object (White \& Ghez 2001),
from which significant photometric variability can be
expected. Photometric variability has indeed been documented
for several IRCs (Ghez et al. 1991, Koresko et al. 1997),
including {\vd} (\S\,\ref{subsec:sed}). Because IRCs are, by
definition, encountered near few Myr-old T\,Tauri stars, it is
difficult to imagine that they recently formed from the
contraction of a vast (several thousand AU) envelope as normal
T\,Tauri stars do, casting some doubt upon their protostellar
nature. However, an isolated object with the same photometric
and spectroscopic properties as {\vd} would unambiguously be
classified as an embedded protostar. The future detection of
photospheric features in the spectrum of {\vd} might clarify
its properties, but for now we cannot conclusively determine
the nature of this object.

Among the seven IRCs studied by Koresko et al. (1997), six
were found to be more massive than their optically bright
``primaries'', which suggested that their peculiarly dusty
environment were a consequence of being the most massive
component in the systems. However, the new IRC discovered here
within {\vt} is found to be the lowest luminosity/mass object
of the quadruple system. We therefore propose that there is a
continuum of IRC masses and luminosities. In other words,
there may be many yet undetected IRCs, as observational
studies of such objects have likely been biased towards the
brightest and therefore most massive ones. If so, we may
expect that further deep, high-angular resolution
10--20\,$\mu$m imaging around known TTS will reveal a
significant population of much fainter objects with SEDs
similar to that of {\vd}.


\subsection{A ``fully mixed'' system}
\label{subsec:ctt}

Most IRCs studied so far are associated with CTTS (Koresko et
al. 1997). However, the quadruple system {\vt} consists of two
1--1.5$\,M_\odot$ WTTS in a SB, a 0.7$\,M_\odot$ CTTS orbiting
around it at $\sim15$--25\,AU, and a low-mass IRC located
further away. This system can therefore be considered as a
``fully mixed'' system, containing all three types of TTS
known so far: CTTS, WTTS and IRC. How it has evolved in the
last Myr or so remains mysterious and it may even seem
puzzling that a multiple system with two WTTS can also contain
more active objects such as a CTTS and an IRC. However, one
can speculate that strong interactions between the various
components and their circumstellar environment have been major
factors in its evolution.

The presence of two close companions located $\sim$0.3\,AU and
$\sim$15\,AU away from {\vt}\,A can naturally explain why the
two components of the SB have rapidly evolved toward WTTS. The
various orbital motions would indeed clear out any
circumstellar disk except in a narrow annulus between
$\sim1$--5\,AU (e.g., Artymowicz \& Lubow 1994) within a few
thousand years. Similarly, the circumstellar disk surrounding
{\vc} would be rapidly truncated to an outer radius of at most
3--4\,AU. While the material surrounding {\vab} appears to
have fully dissipated, {\vc} still possesses an active
accretion disk. However, the lifetime of such a small disk is
relatively short unless it supports only a very small
accretion rate. The upper limit on the millimeter thermal
emission from the system corresponds to a maximum possible
disk mass of only $\sim10^{-2}\,M_\odot$ (Beckwith et
al. 1990; Dutrey et al. 1996). Given the age of the system and
assuming that all this mass is located around {\vc}, only a
very low {\it time-averaged} accretion rate ($<\dot{M}>\
\lesssim 2 \times 10^{-9}\,M_\odot$/yr) could have allowed the
disk to survive for so long. Most likely, this circumstellar
disk is being replenished from an outer reservoir of material,
as has already been suggested for tight pre-main sequence
binaries (e.g., Prato \& Simon 1997). The fact that {\vc}
seems to be the only component of the central triple system to
be replenished may suggest that the orbital motion of the SB
prevents material from falling deep into the potential well of
{\vab} due to resonance interactions with the two stars.

Since it is on a much wider orbit, {\vd} can retain a much
larger envelope, several tens of AU in diameter, and its
dissipation timescale may not be shorter than the age of the
system. Although replenishment is not required to account for
the presence of circumstellar material around this source, it
is nonetheless surrounded by an unusually optically thick
envelope. This may be a transient state generated through
interactions with the other components of the system. In any
case, {\vt} is the first system among T\,Tauri stars to
display such a range of evolutionary states and it is likely
that the compactness of the system, through strong star-star
interactions, has an important role to play in this variety.


\subsection{Do passive disks around TTS really exist?}
\label{subsec:passive}

The 5--60\,$\mu$m flux from {\vt}, that was previously
assigned to a truncated, fossil, circumstellar/circumbinary
disk around the WTTS system {\vab}, is now believed to be
associated with the newly identified IRC, {\vd}, with a
possible contribution of the disk surrounding the CTTS
{\vc}. We conclude that there is no evidence for a fossil disk
in this system.

Candidate fossil disks among pre-main sequence stars are quite
rare. Skrutskie et al. (1990) and more recently Stassun et
al. (2001) obtained 10\,$\mu$m photometry for relatively large
samples (83 TTS and 32 WTTS respectively) and only identified
a handful of such cases among pre-main sequence objects in
Taurus and Orion. The most suggestive candidate is GM\,Aur
which is known to be surrounded by a dusty disk in Keplerian
rotation (Koerner, Sargent \& Beckwith 1993) but shows at best
a marginal photometric excess shortwards of 10$\,\mu$m
(Stassun et al. 2001, White \& Ghez 2001). However, this
object presents a strong accretion-induced ultraviolet/blue
excess and it is unclear whether it should be classified as a
passive disk. About ten other objects, or roughly 10\,\% of
the combined sample, might fall in this category, suggesting a
lifetime for this phenomenon on order of a few $10^5$\,yrs.

However, our analysis of {\vt} has led us to remove this
system from the list of candidate fossil disks and one may
wonder whether other systems will similarly be reclassified
out of this category by further analysis. If so, fossil disks
would be extremely short-lived structures. As we have shown
here, the presence of a close, unresolved IRC in the vicinity
of a WTTS can be misinterpreted as a passive disk on the basis
of its SED. Similarly, a tight SB would rapidly clear a
central gap in its surrounding disk and the absence of the
hottest parts of the disk would result in a dip in the NIR
part of the SED of the system (e.g., Mathieu et al. 1991). A
disk inner radius of about 0.4\,AU results in no excess above
photospheric levels up to 5\,$\mu$m or so, similar to the SED
of a passive disk. Such an inner radius can be created by a
binary with a semi-major axis of about 0.15\,AU (Artymowicz \&
Lubow 1994), corresponding to an orbital period of about
$20\,d$ for a $1\,M_\odot$ total system mass. Unresolved IRCs
and short period SBs can therefore mimic passive disks, and it
is therefore important to compare the frequencies of these
various systems to decide whether a significant proportion of
passive disks really exists among pre-main sequence stars.

So far, at least five IRCs out of $\sim100$ pre-main sequence
objects are known in Taurus (T\,Tau\,Sa, Haro\,6-10\,N,
XZ\,Tau\,B, UY\,Aur\,B and now {\vd}) and more may be
discovered when 5--10\,$\mu$m high-angular images of large
samples of TTS are performed. This observed proportion of
$\sim5\,\%$ of IRCs is therefore only a lower
limit\footnote{IRCs constitute a significant fraction of the
``high accretion rate'' population identified by White \& Ghez
(2001) in Taurus on the basis of the extremely red colors of
these objects ($K-L>1.4$, as is the case for {\vd}). Roughly
10\,\% of all TTS belong to this other category, in
qualitative agreement with our estimated proportion of
IRCs.}. The proportion of SBs with $P_{orb}<20\,d$ for
solar-type main sequence objects is estimated to be about
4--5\,\% (Duquennoy \& Mayor 1991), and it might be even
higher among young stars in Taurus as is the case for visual
binaries (e.g., Ghez et al. 1993). To the best of our
knowledge, only 4 such short-period SBs are known among the
Taurus pre-main sequence population, out of which only 2
(V\,826\,Tau, Mundt et al. 1983, and LkCa\,3, Mathieu 1994)
are classified as WTTS based on their visible/NIR photometric
and spectral properties. Overall, the proportion of passive
disks ($\lesssim10\,\%$) qualitatively matches those of IRCs
and short period SBs among pre-main sequence stars (both
$\gtrsim5\,\%$). It is very possible that most (if not all)
fossil disks candidates are in fact associated with currently
unknown IRCs or SBs. More generally, this emphasizes how
dangerous SED analyses are if all stellar components have not
been resolved. This issue, which was suggested by Chelli et
al. (1988) and Ghez (1996), resurfaces as a concern with the
advent of IR missions with only moderate spatial resolution
($\gtrsim1\arcsec$), such as SIRTF and SOFIA.


\section{Conclusion}
\label{sec:conclu}

Using speckle interferometry techniques at the 10m Keck I and
200-inch Hale telescopes and the AO system on Keck II, we have
obtained multi-epoch high-angular resolution images of the
multiple system {\vt} in the visible and the NIR throughout
the last decade. It had been considered one of the rare
examples of the so-called ``passive'' or ``fossil'' disks
which are thought of as disks in which planetesimals have
formed in the inner parts, thus halting the accretion process
on the stars without expelling the outer disk material.

In addition to the unresolved SB {\vab} and the previously
known speckle companion {\vc}, we identify a $K\sim9$,
$K-L'\sim1.7$\,mag, fourth component in our AO images and in
some speckle datasets. {\vd}, which is located only 0\farcs2
away from {\vab}, is the brightest source in the system at
$4.8\,\mu$m. We have also obtained medium-resolution 2\,$\mu$m
spectra of {\vab} and {\vd}. Given its NIR brightness, its
location, and its apparent motion, we conclude that the newly
identified component is physically associated with the
system. We fit the motion of {\vc} around {\vab} with a closed
Keplerian orbit, and we conclude that the quadruple system can
be dynamically stable provided {\vd} is on a highly inclined
orbit, as are the other orbits within the system. As opposed
to T\,Tau (Loinard et al. 2003), this system does not support
the proposed explanation to the observed overabundance of
young high-order multiple systems that many of them are in
fact unstable systems that will rapidly decay into an ejected
single star and a lower order multiple system (Ghez et
al. 1993; Reipurth 2000).

Analysis of the SED of the system and of our spectra supports
the following picture: both components of {\vab} are
1--1.5$\,M_\odot$ WTTS, {\vc} is a lower-mass variable CTTS
and {\vd} is a new member of the IRC category, i.e. a deeply
embedded companion to a normal T\,Tauri star. The properties
of this new object include a broadly peaked SED from
$\sim5$--20$\,\mu$m, significant variability at 2$\,\mu$m, and
a featureless 2$\,\mu$m spectrum. Its low bolometric
luminosity ($L_D \lesssim 0.8\,L_\odot$) shows that it is the
least massive component of the system, in contrast to most
other IRCs. We suggest that many low-mass IRCs might remain to
be discovered. The physical nature of IRCs remains
unconstrained since both embedded T\,Tauri stars or less
evolved protostars are almost identical in terms of their
observable properties.

The immediate implication of the presence of {\vd} is the
rejection of the hypothesis of a passive disk in the system
since the long known IR excess is now attributed to the
IRC. Given the estimated proportions of IRCs and short period
SBs, we argue that the frequency of passive disks among
pre-main sequence stars may have been strongly overestimated
in the past and that the lifetime of this transition phase may
be much shorter than previsouly thought. More generally, our
analysis of {\vt} shows how cautious one must be when using
NIR or mid-IR photometry of TTS obtained with low spatial
resolution devices.


\acknowledgements

Part of the data presented herein were obtained at the
W. M. Keck Observatory, which is operated as a scientific
partnership among the California Institute of Technology, The
University of California, and the National Aeronautics and
Space Administration. The Observatory was made possible by the
generous financial support of the W. M. Keck Foundation. We
gratefuly thank the staff at Palomar and Keck observatories
for their assistance during the observations as well as Bruce
Macintosh for his help during the acquisition of part of the
data presented in this study. A detailed report by an
anonymous referee significantly helped improve this
manuscript. Frank Marchis kindly provided the synthetic PSFs
that were used in this study.

This work has been supported in part by the National Science
Foundation Science and Technology Center for Adaptive Optics,
managed by the University of California at Santa Cruz under
cooperative agreement AST 98-76783, by NASA's Origins of Solar
System program grant NAG-6975, and by the Packard
Foundation. This research has made use of the SIMBAD database,
operated at CDS, Strasbourg, France. The authors wish to
recognize and acknowledge the very significant cultural role
and reverence that the summit of Mauna Kea has always had
within the indigenous Hawaiian community. We are most
fortunate to have the opportunity to conduct observations from
this mountain.



\clearpage

\begin{deluxetable}{cccccccccc}
\tabletypesize{\scriptsize}
\tablecaption{Summary of all spatially resolved measurements
of the {\vt} system.}
\startdata
\tableline
Date & Obs. & Filter & \multicolumn{3}{c}{{\vab}--C} &
\multicolumn{3}{c}{{\vab}--D} & \\ 
 & Tech. & & $\rho$ (mas) & $\theta$ (\degr) & $\Delta m$ (mag) &
$\rho$ (mas) & $\theta$ (\degr) & $\Delta m$ (mag) &  Ref. \\
\tableline
\tableline
1990 Oct 03 & Spk & $K$ & 112$\pm$2 & 295$\pm$4 & 0.80$\pm$0.03 &
(...) & (...) & (...) & 2 \\
1991 Sep 21 & 1D\tablenotemark{a} & $K$ & 170$\pm$10 & 295$\pm$3 &
2.22$\pm$0.33 & (...) & (...) & (...) & 3 \\ 
1991 Sep 25-29 & Spk & $F_{620}$ & 106$\pm$5 & 298$\pm$2 &
2.94$\pm$0.10 & (...) & (...) & (...) & 1 \\
'' & & $F_{656}$ & '' & '' & 2.19$\pm$0.10 & (...) & (...) & (...) & 1
\\ 
'' & & $F_{800}$ & '' & '' & 2.05$\pm$0.10 & (...) & (...) & (...) & 1
\\ 
'' & & $F_{850}$ & '' & '' & 2.19$\pm$0.10 & (...) & (...) & (...) & 1
\\ 
1992 Oct 10 & Spk & $K$ & 120$\pm$20 & 307$\pm$5 & 1.71$\pm$0.01 &
(...) & (...) & (...) & 4,5 \\ 
1993 Oct 05 & Spk\tablenotemark{b} & $J$ & 111$\pm$4 & 308.9$\pm$1.4 &
2.37$\pm$0.05 & (...) & (...) & (...) & 6 \\ 
1993 Nov 25 & Spk & $K$ & 92$\pm$8 & 304$\pm$8 & 1.75$\pm$0.22 & (...)
& (...) & (...)& 4,5 \\
1994 Oct 19 & Spk & $K$ & 65.6$\pm$1.4 & 318$\pm$2 & 0.46$\pm$0.04 &
(...) & (...) & (...) & 4,5 \\
1994 Oct 29 & $HST$ & $U'$ & 62.8$\pm$2.4 & 321.9$\pm$2.7 &
1.89$\pm$0.78 & (...) & (...) & (...) & 5 \\
'' & '' & $B'$ & '' & '' & 2.78$\pm$0.92 & (...) & (...) & (...) & 5
\\ 
'' & '' & $V'$ & '' & '' & 1.73$\pm$0.44 & (...) & (...) & (...) & 5
\\ 
'' & '' & $R'$ & '' & '' & 1.36$\pm$0.37 & (...) & (...) & (...) & 5
\\ 
'' & '' & $I'$ & '' & '' & 1.19$\pm$0.54 & (...) & (...) & (...) & 5
\\ 
1996 Oct 24 & Spk\tablenotemark{c} & $K$ & (...) & (...) &
(...) & 162$\pm$29 & 176$\pm$5 & 2.17$\pm$0.27 & 1 \\ 
1997 Nov 19 & Spk\tablenotemark{c} & $K$ & (...) & (...) &
(...) & 187.7$\pm$3.9 & 173.7$\pm$1.5 & 2.12$\pm$0.10 & 1 \\
1998 Nov 04 & Spk & $K$ & 57.4$\pm$1.2 & 67.1$\pm$1.4 & 0.76$\pm$0.03
& 191.3$\pm$3.6 & 164.7$\pm$1.3 & 2.85$\pm$0.06 & 1 \\
1999 Nov 20 & Spk & $K$ & 74.4$\pm$1.4 & 81.9$\pm$1.1 & 0.83$\pm$0.06
& 202.4$\pm$3.8 & 161.3$\pm$2.0 & 2.79$\pm$0.05 & 1 \\
2000 Nov 20 & AO & $H$ & 85$\pm$3 & 88.4$\pm$1.5 & 0.44$\pm$0.10 & 
212$\pm$4 & 155.7$\pm$0.7 & 2.86$\pm$0.15 & 1 \\
'' & '' & $K$ & '' & '' & 0.46$\pm$0.05 & '' & '' & 1.92$\pm$0.10 & 1
\\  
2000 Dec 08 & Spk & $K$ & 91.1$\pm$3.9 & 88.5$\pm$1.2 & 1.01$\pm$0.07 &
217.4$\pm$4.9 & 159.3$\pm$4.9 & 2.43$\pm$0.07 & 1 \\
2001 Feb 08 & Spk\tablenotemark{d} & $K$ & 95$\pm$5 &
93.5$\pm$0.6 & (...) & (...) & (...) & (...) & 7 \\
2001 Nov 02 & Spk\tablenotemark{d} & $K$ & 99$\pm$7 &
94.6$\pm$1.0 & (...) & (...) & (...) & (...) & 7 \\
2001 Dec 08 & Spk & $K$ & 100.0$\pm$1.0 & 92.3$\pm$1.0 & 0.33$\pm$0.07
& 223.0$\pm$3.0 & 154.1$\pm$1.0 & 1.84$\pm$0.07 & 1 \\
2002 Dec 13 & AO & $L'$ & 116.5$\pm$2.0 & 100.6$\pm$0.7 &
0.06$\pm$0.05 & 237$\pm$3 & 153.7$\pm$0.6 & 0.36$\pm$0.05 &
1\\
'' & '' & $M'$ & '' & '' & -0.21$\pm$0.05 & '' & '' &
-0.30$\pm$0.05 & 1\\
\enddata
\tablecomments{The data presented in this table have been obtained
through two-dimensional (``Spk'') and one-dimensional (``1D'')
speckle interferometry, AO or $HST$/WFPC2 imaging. $\rho$
represents the binary separation and $\theta$ its position
angle as measured East from North on the sky. In all
observations between 1990 and 1997, only one companion was
unambiguously detected.}
\tablenotetext{a}{This measurement was obtained by analyzing two
orthogonal one-dimension scans and retrieving the actual
separation and position angle by deprojection from these
axes. The uncertainties associated with this method are
significantly larger from those associated with two-dimension
speckle measurements, and this data point differs by more than
5\,$\sigma$ from the visible 2D speckle result we obtained
just a week later, suggesting that the uncertainties of the 1D
measurement are underestimated. It has not been included in
our final analysis.}
\tablenotetext{b}{Instead of being a Nyquist-sampled
measurement, this dataset was obtained with a pixel scale that
is almost equal to the diffraction limit of the 3.5\,m Calar
Alto telescope in the $J$ band (0\farcs07/pixel and 0\farcs074
respectively). This measurement is thus significantly aliased
and therefore has not been used in our analysis.}.
\tablenotetext{c}{In these measurements, the tight pair
{\vab}--C was not fully resolved. All photometric and
astrometric information extracted from this dataset therefore
pertains to the {\vab}C--D system.}
\tablenotetext{d}{These measurements were obtained with a
different instrumental and calibration set-up than most of our
data. To minimize the effect of systematic differences between
datasets, we do not include these datasets in our analysis
since our own data already provide an adequate time sampling
throughout the 1998--2002 period.}
\tablerefs{1) This work; 2) Ghez et al. (1993); 3) Leinert et
al. (1993); 4) Ghez et al. (1995); 5) Ghez et al. (1997); 6)
Woitas et al. (2001); 7) Tamazian et al. (2002).}
\end{deluxetable}
\clearpage

\begin{deluxetable}{cccccccc}
\tablecaption{Parameters for the orbit of {\vc} around {\vab}.}
\startdata
\tableline
$M_{ABC} (M_\odot)$ & $P_{orb}$ (yrs) & $a$ (AU) & $e$ & $i$
(\degr) & $T_0$ & $\omega$ (\degr) & $\Omega$ (\degr) \\
\tableline
$3.7\pm0.7$\tablenotemark{a} & $46\pm6$ & $20\pm1$ &
$0.3\pm0.1$ & $66\pm3$\tablenotemark{b} & $1996.5\pm0.8$ &
$81\pm10$\tablenotemark{b} & $288\pm1$\tablenotemark{b} \\
\enddata
\tablenotetext{a}{The uncertainty on the total system mass
only includes the fitting uncertainty. An additional 10\,\%
random uncertainty arises from the error on the distance to
the system since orbital analysis relying on astrometric data
only constrains $M_{syst}/D^3$.}
\tablenotetext{b}{In the absence of radial velocity
measurements, an ambiguity in the inclination angle remains
unsolvable. We have {\it assumed} that $i<90$\degr, which
means that {\vc} is currently in front of the other
component. However, it is equally possible that $i=114$\degr,
in which case $\omega=279$\degr\ and $\Omega=108$\degr. The
uncertainties on these parameters, as well as the location of
periastron, are however unaffected by this ambiguity.}
\end{deluxetable}
\clearpage

\end{document}